\documentclass[a4,oldversion]{aa}
\usepackage{txfonts,graphicx}
\usepackage{amssymb}
\usepackage{natbib}
\newcommand{\mapiii}{MAPPINGS  \textsc{iii}}
\newcommand{\hi}{H\,{\sc i}}
\newcommand{\hii}{H\,{\sc ii}}
\newcommand{\oiii}{O\,{\sc iii}}
\newcommand{\silii}{Si\,{\sc ii}}
\newcommand{\sii}{S\,{\sc ii}}
\newcommand{\siii}{S\,{\sc ii}}
\newcommand{\neii}{Ne\,{\sc ii}}
\newcommand{\neiii}{Ne\,{\sc iii}}
\newcommand{\nev}{Ne\,{\sc v}}
\newcommand{\oiv}{O\,{\sc iv}}

\def\phn{\phantom{0}}

\def\phs{\phantom{$-$}}

\catcode`\@=11
\def\gapprox{\mathrel{\mathpalette\@versim>}}
\def\lapprox{\mathrel{\mathpalette\@versim<}}
\def\@versim#1#2{\lower2.45pt\vbox{\baselineskip0pt\lineskip0.9pt
      \ialign{$\m@th#1\hfil##\hfil$\crcr#2\crcr\sim\crcr}}}
\catcode`\@=12
\newcommand{\mum}{\ensuremath{\mu\mbox{m}}}

\newcommand{\pccm}{\ensuremath{\,\mbox{cm}^{-3}}}
\newcommand{\pscm}{\ensuremath{\,\mbox{cm}^{-2}}}

\newcommand{\flux}{\ensuremath{\,\mbox{erg}\,\mbox{cm}^{-2}\mbox{s}^{-1}}}

\begin{document}
  \title{The Infrared Emission from the \\Narrow Line Region}
  \titlerunning{The IR Emission from the NLR}
  \author{Brent Groves\inst{1}, Michael Dopita\inst{2}, \& Ralph 
Sutherland\inst{2}}
  \authorrunning{B.~Groves, M.~Dopita, \& R.~Sutherland\inst{2}}

  \institute{Max Planck Institute for Astrophysics,\\
             Karl-Schwarzschild str. 1, Garching, 85741
             Germany
             \\
             \email{brent@mpa-garching.mpg.de}
\and Research School of Astronomy \& Astrophysics,
The Australian National University, \\
Cotter Road, Weston Creek, ACT 2611, Australia
                 }
\date{Received <date> / Accepted <date>}
\abstract{
We present models for the mid- and far- infrared emission 
from the Narrow Line Region (NLR) of Active Galactic Nuclei (AGN).
Using the \mapiii\ code we explore the effect of typical NLR 
parameters on the spectral characteristics of the IR emission. These 
include useful IR emission line ratio diagnostic diagrams for the 
determination of these parameters, as well as Star formation--AGN
mixing diagnostics. We also examine emission line to continuum 
correlations which would assist in separating the IR emission 
arising from the NLR from that coming from the inner torus. 
We find for
AGN like NGC 1068 and NGC 4151 that the NLR only contributes
$\sim$10\% to the total IRAS 25 \mum\ flux, and that other components
such as a dusty torus are necessary to explain the total AGN  IR emission.}
{}

\maketitle
\section{Introduction}
Active galactic nuclei (AGN) are well known to emit strongly at all 
wavelengths, from gamma-rays and X-rays to the infrared and radio. 
Their emission characteristics at each wavelength gives us insight 
into the central engine of these galaxies, whether observed directly, 
such as at X-ray wavelengths, or indirectly through reprocessed 
emission, such as by study of the emission lines from the narrow line 
region (NLR).

One area that has seen recent rapid advances is 
studies of the mid-- and far--infrared characteristics of AGN, with 
space telescopes such as IRAS, ISO and Spitzer making this wavelength 
regime readily accessible. As the continuum in this wavelength 
region is dominated by  warm and hot  dust re-emission, it is 
particularly useful in testing the paradigm of the unified  model,
which argues for a dense equatorial band of dust surrounding the 
black hole central engine \citep{Antonucci85}.

Various attempts have been made to model the mid-- and far--IR 
emission.  These range from simple empirical fits of dust grey-body 
emission to the observations \citep{Edelson86,Blain03}, to more 
complicated distributions of dust which include the stochastic heating of 
dust by an AGN power-law continuum \citep{Dullemond05}. With the 
prevalence of the unified model, most analyses of the IR from AGN 
have concentrated on the emission from dust lying in a dense torus 
surrounding a nuclear
heating source \citep{Pier92,Pier93,Granato94,Efstathiou95, 
Granato97,vanBemmel04,Nenkova02}. However, there have been some 
authors who have considered different geometries 
\citep[i.e.][]{Siebenmorgen04} or have considered in more detail the 
collisional heating and destruction of dust \citep{Contini04}.

Often ignored in the considerations of AGN IR emission is the
contribution from the narrow line 
region (NLR). This region is known to contain dust 
\citep[e.g.][]{Radomski03} and, 
like the dusty torus, will be heated by the central engine and 
re-emit at IR wavelengths \citep{Netzer93}. With 
lower densities than found in the central torus, the NLR  also emits strong 
forbidden lines  such as the  [\oiii] 500.7 nm line. Such lines are 
often one of the main diagnostics for the presence of an AGN, and 
also provide important constraints on the form and intensity of 
the radiation field which excites them. They also provide insight 
into the physical conditions such as density or pressure, and 
constraints on the metallicity of the circum-nuclear ISM. 
 
As AGN are often heavily dust-embedded, much of the line emission from 
the near nuclear parts of the NLR may be totally obscured at optical 
wavelengths. To penetrate this circum-nuclear dust while retaining 
adequate spatial resolution to investigate the structure of the dense 
parts of the NLR, we need to observe at mid-IR wavelengths. 
Instruments such as the \emph{IRS} instrument of the 
\emph{Spitzer Space Telescope} are finally providing the data needed 
to study this region. In order to interpret this data we need both 
line and continuum diagnostics in the mid-IR. 

Here,  we will 
explore the expected characteristics of the infrared emission from 
the NLR, examine the dependence of this emission on the parameters 
that define the NLR, and investigate possible diagnostics that can be 
used to define these parameters and so give us insight into the NLR 
and
the AGN as a whole.

\section{Modelling the Dust in Narrow Line Regions}

The effects of dust on the structure of the NLR and its influence of 
the emission line spectrum has been discussed in detail in previous 
papers \citep{Groves04a,Groves04b}. Here we extend this work using a 
more realistic distribution of dust and considering in addition the 
stochastic heating of the dust to investigate the nature of the 
far-IR dust re-emission spectrum and its relationship to the line 
spectrum of the NLR.

The existence of dust around AGN is well accepted. Indeed, it forms an 
integral component of the unification scenario of active galaxies 
\citep{Antonucci92}. That dust specifically exists within the NLR of 
AGN is also suggested by theoretical models \citep{Netzer93,Dopita02}
and is well supported by recent high resolution IR observations of nearby AGN
\citep{Alloin00,Bock00,Packham05,Galliano05,Mason06}. 

Both the composition and size distribution of the dust in the NLR are 
still uncertain. Current observations \citep{Gaskell04} suggest that 
the properties of dust in the vicinity of AGN may differ to that 
found in our Galaxy. This would be unsurprising considering the 
gas pressure, UV radiation field and dynamics of this environment are 
all much more extreme than those found in the solar vicinity. 
Polycyclic aromatic hydrocarbons (PAHs), while common in our Galaxy 
and are observed in starforming galaxies, are not expected in the 
excited regions of active galaxies. This is because of the strong EUV and X-ray
emission from the AGN, which are thought to effectively destroy PAHs though 
photo-dissociation \citep{Voit91,Voit92}. In the cases where PAH
emission is seen in the nuclear regions of AGN, the emission is
thought to arise from nuclear starbursts. The starbursts supply the
Far-UV needed to excite the PAH emission, but would be shielded from the X-ray
emission of the AGN by intervening dusty, obscuring material
\citep{Imanishi03,Imanishi04}.

The lack of these planar 
linked benzene-ring molecules may also extend to small grains, with
the smallest graphite and silicate grains possibly being similarly destroyed by
the strong radiation field.
As first suggested by 
\citet{Laor93}, the idea that larger grains dominate the dust in AGN 
relative to that found in the Galaxy has been supported by 
observations such as the recent work of  \citet{Gaskell04}, which 
demonstrated the much flatter attenuation curves in AGN in the UV. 
\citet{Maiolino01a,Maiolino01b} explored the observational evidence 
for different dust properties within  AGN and came to the conclusion 
that the most favorable scenario was one where large dust grains 
dominated the distribution of AGN dust.
However, flatter attenuation curves may also arise in a highly clumpy 
dust distribution \citep{Fischera04,Fischera05}. 

The composition of dust within AGN is also poorly constrained by 
observation. Metals appear to be depleted within the NLR gas, 
although the magnitude of this depletion is uncertain. In terms of 
extinction, the standard silicate + graphite dust appears to 
reproduce the observations reasonably well 
\citep{Laor93,Maiolino01a,Maiolino01b}, although additional dust 
components, such as diamond dust have also been suggested 
\citep{Binette05}.

Using this information  we have tried to construct a simple yet 
physically plausible model for the dust in the NLR clouds. We assume 
two types of dust: graphitic dust and silicaceous dust. The optical 
data for these are taken from work of \citet{Draine84}, 
\citet{Laor93} and \citet{Weingartner01}.

The grain size distribution of both types of dust is taken to be a 
power-law, with the index arising from grain shattering. Grain 
shattering has been shown to lead naturally to the formation of a 
power-law size distribution of grains with index $\alpha \sim 3.3$ 
\citep{Jones96}. This is within the bounds of the standard 
\cite{MRN77} value of $\alpha \sim 3.5$. Such a grain size 
distribution can also hold only between certain limits in size, the 
minimum size being determined by natural destruction processes such 
as photodestruction, and the upper limit being determined by limits 
on the growth by condensation and sticking. These may occur within 
the AGN or previous to the dusty gas entering the energetic region. 
To capture these elements of the physics, we have adopted a modified 
grain shattering profile with the form;
\begin{equation}
dN(a)/da = k  a^\alpha
\frac{e^{-(a/a_\mathrm{min})^{-3}}}{1+e^{(a/a_\mathrm{max})^3}}. \label{GSprof}
\end{equation}

This creates a smooth exponential cut-off in terms of the grain mass 
at both the minimum and maximum grain size of the distribution. This 
smooth cut removes any edge effects in either the emission or 
extinction by dust that arise due to the sharp cut-offs in the other 
distributions. To implement the larger grain dominated distribution 
discussed above, we have assumed that the smallest grains have been 
destroyed due to the energetic environment, and used a larger minimum 
grain size relative to Galactic of $a_{\rm min} = 0.01$ \mum.

The constant k is determined by the total dust-to-gas mass ratio, 
which is determined by the depletion of the heavy elements onto dust. 
Our models use a solar abundance set (as modified by the latest 
abundance determinations 
\citet{Allende01,Allende02,Asplund00a,Asplund00b}). The depletion 
factors are those used by \citet{Dopita00} for starburst and active 
galaxy photoionization modeling and are similar to those found by 
\citet{Jenkins87} and \citet{SavageS96} in the local ISM using the UV 
absorption lines to probe various local lines of sight.

Within the \mapiii r code, the dust grain size distribution is 
divided into 80 bins spaced logarithmically between 0.001--10 \mum. 
The number of grains of each type in each bin is then determined by 
equation \ref{GSprof}. The absorption, scattering and photoelectric 
heating is calculated for each size bin. This is then used to 
establish the temperature probability distribution for computation of 
the FIR re-emission spectrum with quantum fluctuations.

\section{Calculating the IR Emission from Dust}

In energetically active regions such as found around starbursts and 
AGN it is essential to take stochastic quantum heating of dust grains 
properly into account. In particular, the temperature fluctuations of 
grains serves to provide a population of small grains which are 
hotter than their equilibrium temperature and thus which show an 
excess in their mid-IR emission.

Temperature fluctuations and stochastic heating processes have been 
considered by several authors since Greenberg first proposed these 
ideas \citep[e.g.][]{Duley73,Greenberg74,Purcell76,Dwek86}.  The most 
detailed work on dust and temperature fluctuations has been a series 
of papers by Draine \& co--authors \citep{Draine85,Guhatha89} 
culminating in a set of papers with Li \citep{Draine01,Li01}.

The FIR emission treatment used in the \mapiii r code is based on the 
algorithms of \citet{Guhatha89} (GD89) and \citet{Draine01} (DL01). 
The code provides an optimized solution of their algorithms to 
determine the dust grain temperature distributions for each grain 
size according to the both the strength and detailed spectrum of the 
local radiation field. The code then integrates the resultant FIR 
emission of the ensemble of dust grains to provide the local 
re-emission spectrum, which is itself integrated through the model in 
the outward-only approximation \citep{Brent04}.

In general, the average temperature of dust grains decreases from the
front of a photoionized cloud exposed to the radiation to the back as
expected. However the average dust temperature is not an exact
representation of the grains, with the largest grains at lower
temperatures, then the temperature of the grains increasing with
decreasing grain size. At a certain size stochastic heating becomes
important and the probability distribution of the dust peaks at a
temperature below the total average dust temperature, but with a tail
extending to very high temperatures.

\section{Narrow Line Region Models}

The physical details of the dusty, radiation-pressure dominated 
photoionization models used here to model the IR continuum of the 
narrow line region clouds have been discussed in depth in previous 
papers \citep{Groves04a,Groves04b}. Here we only provide a brief 
overview of the different input parameters and the parameter space 
explored.

\subsection{Input Ionizing Spectrum}

A simple power-law radiation field, with an index around $\sim -1.4$ 
and extending from the FUV to X-ray, is able to reproduce the 
dominant features observed in NLRs \citep[see e.g.][]{Ost89}. 
However, when the full spectral energy distribution of the NLR is 
being considered a more physical representation of the ionizing 
spectrum is required. The main problem here is that the intrinsic 
spectrum emitted by the NLR is generally heavily obscured by the 
circum-nuclear interstellar matter in the active galaxy before it 
reaches the observer. This is especially true in the extreme-UV.

Our understanding of  this region of the spectrum relies 
predominantly on photoionization theory and on the theory of 
accretion disk models \citep[see eg.][]{Alexander99,Alexander00}. 
This method gives insight
into the spectrum, but is somewhat dependent upon the modelling used
and the other model parameters. 

However, recent work from \citet{Scott04} and 
\citet{Shang05} looked at QSOs beyond the Lyman limit using space 
telescopes, exploring not only the peak of the
big blue bump of the QSO spectra but also the slope beyond this into 
the EUV. These new observations help to constrain the general shape 
of the AGN spectrum and thus the photoionization and accretion disk 
models.

\begin{figure}[htp]
\includegraphics[width=\hsize]{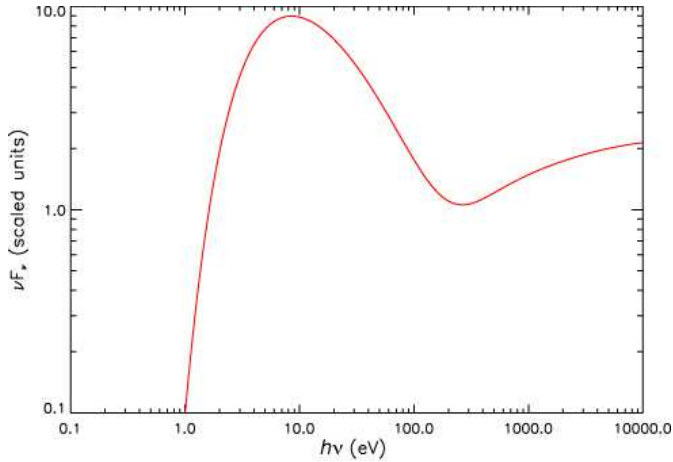}
\caption{The input model ionizing spectrum (scaled) from equation
\ref{eqn:AGNspec}.
}\label{fig:AGNspec}
\end{figure}

We have chosen here to base our input ionizing spectrum predominantly 
on the observations of \citet{Elvis94}, while still taking note of 
the \citet{Scott04} and \citet{Shang05} data. To fit the data we use 
a combination of two power-laws with exponential cut-offs 
\citep{Nagao01,Ferland96},
\begin{equation}\label{eqn:AGNspec}
\begin{array}{rcl}
f_\nu &=& \nu^{\alpha_\mathrm{EUV}}
\exp\left(-\frac{h\nu}{kT_\mathrm{UV}}\right)
\exp\left(-\frac{kT_\mathrm{BBB}}{h\nu}\right)\\
&+& a\nu^{\alpha_\mathrm{X}}
\exp\left(-\frac{h\nu}{kT_\mathrm{X}}\right)
\exp\left(-\frac{kT_\mathrm{BBB}}{h\nu}\right).
\end{array}
\end{equation}
The parameters are chosen to give the most  plausible fit, 
concentrating on the ionizing
spectrum. The final spectrum, shown in figure \ref{fig:AGNspec},  is 
similar to the 3-part broken
power-law often used in photoionization models. The first parameter, 
$\alpha_\mathrm{EUV}$, is defined as the EUV power-law index, and 
hence controls the dominant part of the spectrum. It is
chosen here to be $\alpha_\mathrm{EUV}=-1.75$, which gives a smooth 
fit from the peak of the big blue bump (BBB) and the soft X-ray. This 
fit is made smooth by the UV cut-off, with $kT_\mathrm{UV}=120$ eV. 
The parameter $kT_{\mathrm{BBB}}$ defines the peak of the BBB, and is 
set to $kT_{\mathrm{BBB}}= 7.0$ eV, which gives a peak around 110.0 
nm, as found by \citet{Shang05}.

The second component of equation \ref{eqn:AGNspec} represents the 
X-ray part of the AGN spectrum, which has an index of 
$\alpha_\mathrm{X}=-0.85$ and an upper cut-off simply chosen to
be $kT_\mathrm{X}=10^5$ eV to prevent errors. The parameter $a$ sets 
the scaling between the two components of the AGN spectrum, and is 
set to $a=0.0055$, which gives not only a good fit to the 
\citet{Elvis94} data but also gives an optical-X-ray 
slope\footnote{Defined as in equation 1 from \citet{Nagao01} } of 
index $\alpha_\mathrm{O-X}\sim-1.4$, fitting in with previous 
photoionization models.

\subsection{Narrow Line Region Metallicity and Dust Depletion}

In previous work \citep{Groves04a,Groves04b} we found that the 
observed line ratios for NLR are consistent with a heavy element 
abundance of between solar and twice solar metallicity. Here, for
simplicity, we compute models with solar metallicity, using the 
abundances from \citet[][with references therein]{Dopita05}. For the 
depletion factors, we follow our previous work and use those found in 
the local ISM due to our ignorance about what these factors are 
within the AGN environment.  The solar abundances and depletion 
factors used are given in table
\ref{tab:elements}.

\begin{table}
\begin{center}
\caption{Solar abundance \& metallicity scaling \label{tab:elements}}
\begin{tabular}{lll}
\hline
Element & Abundance\footnotemark &
Depletion\footnotemark\\
\hline
\hline
H .................................. & \phs0.000\phn   & \phs0.00\phn\\
He.................................. & $-$0.987\phn    & \phs0.00\phn\\
C .................................. & $-$3.61\phn\phn & $-$0.30\phn \\
N .................................. & $-$4.20\phn\phn & $-$0.22\phn \\
O .................................. & $-$3.31\phn\phn & $-$0.22\phn \\
Ne.................................. & $-$3.92\phn\phn & \phs0.00\phn\\
Na.................................. & $-$5.68\phn\phn & $-$0.60\phn \\
Mg.................................  & $-$4.42\phn\phn & $-$0.70\phn \\
Al.................................. & $-$5.51\phn\phn & $-$1.70\phn \\
Si...................................& $-$4.49\phn\phn & $-$1.00\phn \\
S .................................. & $-$4.80\phn\phn & \phs0.00\phn\\
Cl.................................. & $-$6.72\phn\phn & $-$0.30\phn \\
Ar.................................  & $-$5.60\phn\phn & \phs0.00\phn\\
Ca.................................  & $-$5.65\phn\phn & $-$2.52\phn \\
Fe.................................. & $-$4.54\phn\phn & $-$2.00\phn \\
Ni.................................. & $-$5.75\phn\phn & $-$1.40\phn \\
\hline
\end{tabular}
\end{center}
\end{table}
\footnotetext[2]{All abundances are logarithmic with
respect to Hydrogen}
\footnotetext[3]{Depletion given as
$\log(X/\mathrm{H})_{\mathrm{gas}}-
\log(X/\mathrm{H})_{\mathrm{ISM}}$}

\subsection{Choice of Variables}

The density and ionization structure of dusty, radiation-pressure 
dominated photoionization models used to represent the NLR has been 
described in detail in \citet{Groves04a}. Here, we follow this
prescription in most respects. However, we here define the initial 
parameters in a slightly different manner in order to provide the 
clearest distinction in distinguishing the effects of each parameter 
upon the infrared emission.

As discussed in the previous work, it is the total pressure of the 
system which defines the density structure of the ionized plasma. The 
total pressure is the sum of both the local gas pressure and the 
incident radiation pressure and can be set as a parameter through 
setting both of these quantities. In a static cloud, assumed here, 
the (negative) gradient of the local radiation pressure must match 
the (positive) gradient of the gas pressure. The effective radiation 
pressure imparted as gas pressure at any point in the cloud depends 
upon both the local ionizing spectrum (both shape and flux density) and 
the optical depth (a function of both the column density and the 
opacity of the gas). As our 
models have the same ionizing spectral shape, the only parameter at a
given optical depth is the incident flux density of the ionizing
radiation.
 
Thus at a given optical depth, such as near the ionization front
(where the dominant ionizing radiation has been absorbed), we have:
\begin{equation}
\begin{array}{rcl}
P_{\rm tot}&=&P_{\rm gas}+P_{\rm rad}\\
&=&P_{0 {\rm gas}}+aI_{0}.
\end{array}
\end{equation}
$P_{0 {\rm gas}}$ is the initial gas pressure of the NLR cloud, $a$ 
is a constant which includes the effect of spectral shape and gas and 
dust opacity, and $I_{0}$ is the incident ionizing flux density upon the 
NLR cloud. 

This results in the same density structure as shown in 
\citet{Dopita02} and \citet{Groves04a}, however with a much better 
defined input. The definition of density structure is important as it
allows us to distinguish which variations in the IR emission arise
from changes in the 
ionizing flux or the NLR density structure.

\subsection{Parameter Space of Models}

Within this work we explore three different total pressures, $P_{\rm 
tot}/k\simeq 10^6$, $10^7$, and $10^8$K\pccm, 
which simulate [\sii]
densities of approximately $10^2, 10^3,$ and $10^4$ \pccm\ 
respectively.

 For each pressure 
regime we have investigated a range of ionizing flux densities. This
range has been chosen to give a reasonable representation of the
typical ionization conditions found within the NLR.  
The values of the input flux densities and corresponding initial gas
pressures are given in table \ref{tab:U}. The total flux density has been
normalised by the factor 3.8098($(P_{\rm tot}/k)/10^6)$\flux\ (or for 
ionizing flux density by 2.0864($(P_{\rm tot}/k)/10^6)$\flux). This
factor results in a ionizing photon density centered around
$S_{*}/c=1$ for the 
$P_{\rm tot}/k =10^6$ models, giving a range of ionization parameters 
centered approximately around $10^{-2}$.

Traditionally, the ionization parameter $U$\footnote{The ionization
parameter $U$ is a dimensionless measure of the number density of ionizing
photons ($S_{*}$), over the gas density; $U=S_{*}/n_{\rm H}c$.} has
been used to describe 
photoionization equilibrium conditions. However in isobaric systems
this parameter is harder to define. So we present here two
parameters for comparison. The
standard ionization parameter as a reference to
previous work, and the $\Xi$ parameter,
\begin{equation}
\Xi=\frac{P_{\rm rad}}{P_{\rm gas}}\simeq\frac{I_{\rm ion}}{nkTc},
\end{equation}
a more suitable parameter to describe the ionization equilibrium in
isobaric systems \citep{Krolik81}. To convert $\Xi_{\rm ion}$ to
include the total incident flux density rather than just the ionizing flux,
multiply by 1.826.

As the 
initial density is uncertain in isobaric models, we define the 
ionization parameter in terms of a normalised initial ionization 
parameter, $\tilde{U}_{0}$. This parameter assumes a 
temperature of $10^4$K to 
obtain the initial density from the input gas pressure. As the 
initial temperature usually exceeds $10^4$K, and the initial 
temperature increases with ionization parameter,  the normalised 
initial ionization parameter underestimates the true initial 
ionization parameter. This becomes increasingly so for higher values. 
The value of both $\tilde{U}_{0}$ and $\Xi_{0}$ 
for each model are given in table 
\ref{tab:U}.  
 
\begin{table}
\caption{Input parameters for NLR models. The total incident flux
density $I_{0}$ is scaled by 3.8098($(P_{\rm tot}/k)/10^6)$\flux\ (see
text), while $P_{0 {\rm gas}}/k$ is
scaled by
$0.1(P_{\rm tot}/k)$}\label{tab:U}
\begin{center}
\begin{tabular}{|c|c|c|c|}
\hline
$I_{0}$ & $P_{0 {\rm gas}}/k$ & $\Xi_{0}$ & $\log (\tilde{U}_{0})$ \\
\hline
4.0 & 1.00 & 20.16    & -0.40 \\
3.0 & 3.20 & \phn4.73 & -1.03 \\
2.0 & 5.40 & \phn1.87 & -1.43 \\
1.0 & 7.60 & \phn0.66 & -1.88 \\
0.5 & 8.70 & \phn0.29 & -2.24 \\
0.25& 9.25 & \phn0.14 & -2.57 \\
0.1 & 9.58 & \phn0.05 & -2.98 \\
\hline
\end{tabular}
\end{center}
\end{table}

The models are truncated at a column density of $\log 
(N$(\hi)$)=21.5$, a reasonable estimate for the NLR clouds 
\citep[e.g.][]{Crenshaw03}, which means that we are comparing like 
absorbing columns when determining the IR emission.

\section{Narrow Line Region SEDs}

\begin{figure*}[htp]
\includegraphics[width=\hsize]{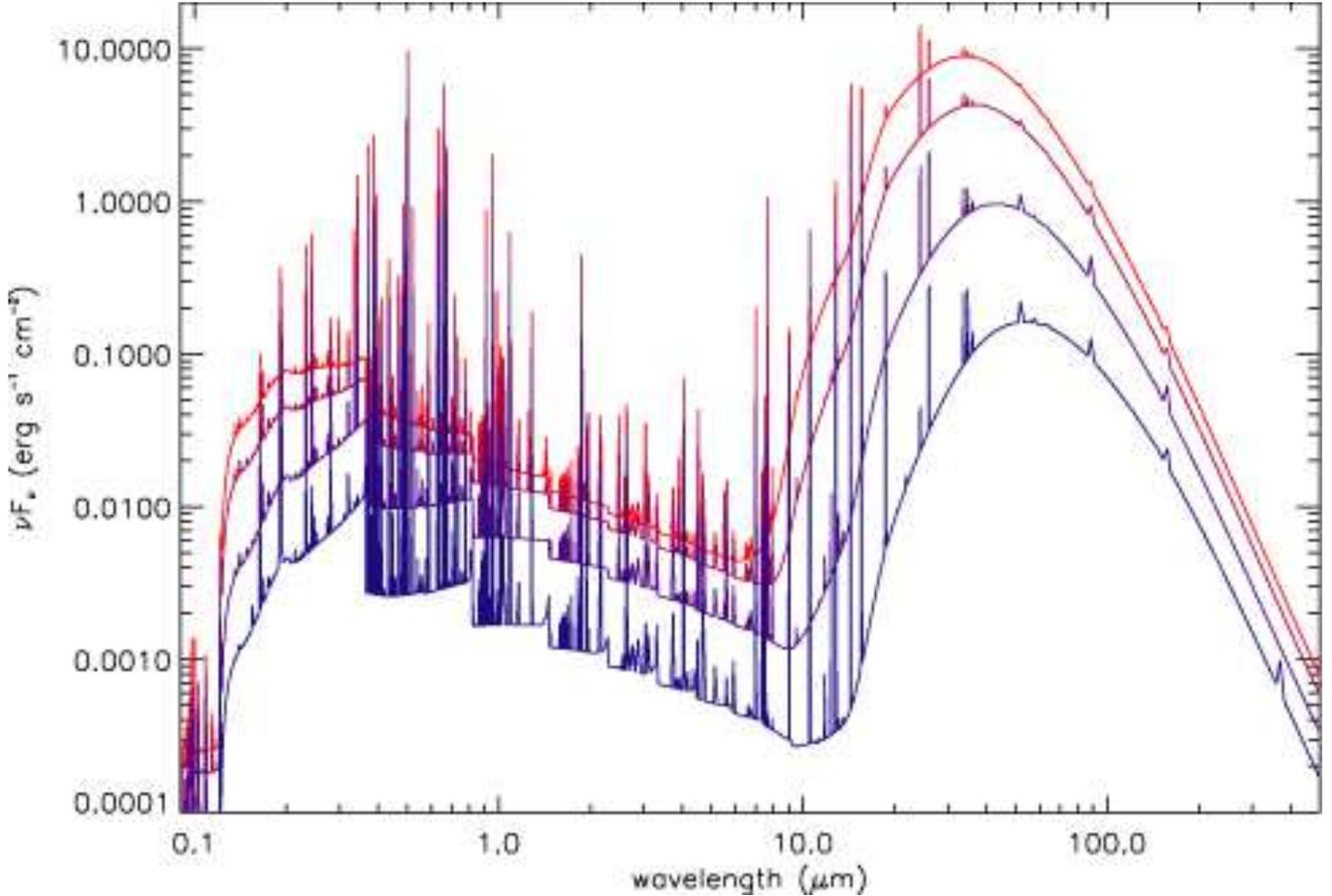}
\caption{Spectral energy distributions (SEDs) for NLR models with 
$P_{\rm tot}/k\simeq10^6$ K \pccm. The models decrease in 
ionization parameter from top to bottom, with $\log
(\tilde{U}_{0})=-0.4,-1.4,-2.2,-3.0$ respectively. 
The SEDs presented here are purely nebular 
emission from the NLR. To obtain the full SED of the AGN, we would need 
to add these SEDs to the attenuated source SED, including any 
contributions from a Broad-Line Region (BLR) and/or accretion 
disk.}\label{fig:1e7}
\end{figure*}

The resulting Spectral Energy Distributions (SEDs) for the $\simeq
10^7$ model are 
shown in figure \ref{fig:1e7}. The 
figure shows four SEDs corresponding to
$\Xi=20.16,1.87,0.29,0.05$.

In this figure we see several features which 
arise due to the combination of the changing ionizing
flux density and the changing ionization parameter. The offset 
between the SEDs arises due to the difference in input flux density. 
A greater incident ionizing flux density results in a greater 
nebular suface brightness.  However, the correlation between the 
resulting nebula surface brightness or emission measure and incident
ionizing flux density is not one to one, due to the varying ratios of 
dust absorbtion compared with absorption in the gaseous phase. 
 
The ionization parameter is one of the main determinants of 
a nebula emission spectrum
\citep[see e.g.][]{ADU03}, and also relates to the dust dominance of 
ionizing photon absorption \citep{Dopita02}. This parameter 
has a strong effect on the overall shape of the final nebula spectrum.

The effect of this parameter on the gaseous emission can be seen, 
in particular, in the stronger \hi\ (eg.~Balmer) edges as we move to lower 
ionization parameters. This is predominantly due to the lower 
electron temperatures which characterize the models with lower 
ionization parameter. The effect of dust can also be seen in several 
ways. For example, the greater overall extinction by dust, the greater 
flux in the UV arising from the input continuum scattered by dust, 
and the intensity of the dust far-IR emission bump. 

As we increase 
in ionization parameter, the fraction of UV flux absorbed by dust 
increases, and this combined with the increase in the ionizing 
intensity results in a greater total of IR emission by dust. It also 
results in hotter dust on average, which means the peak of the IR 
feature is shifted to shorter wavelengths and the flux in the mid-IR 
is greater.

\begin{figure*}[htp]
\includegraphics[width=\hsize]{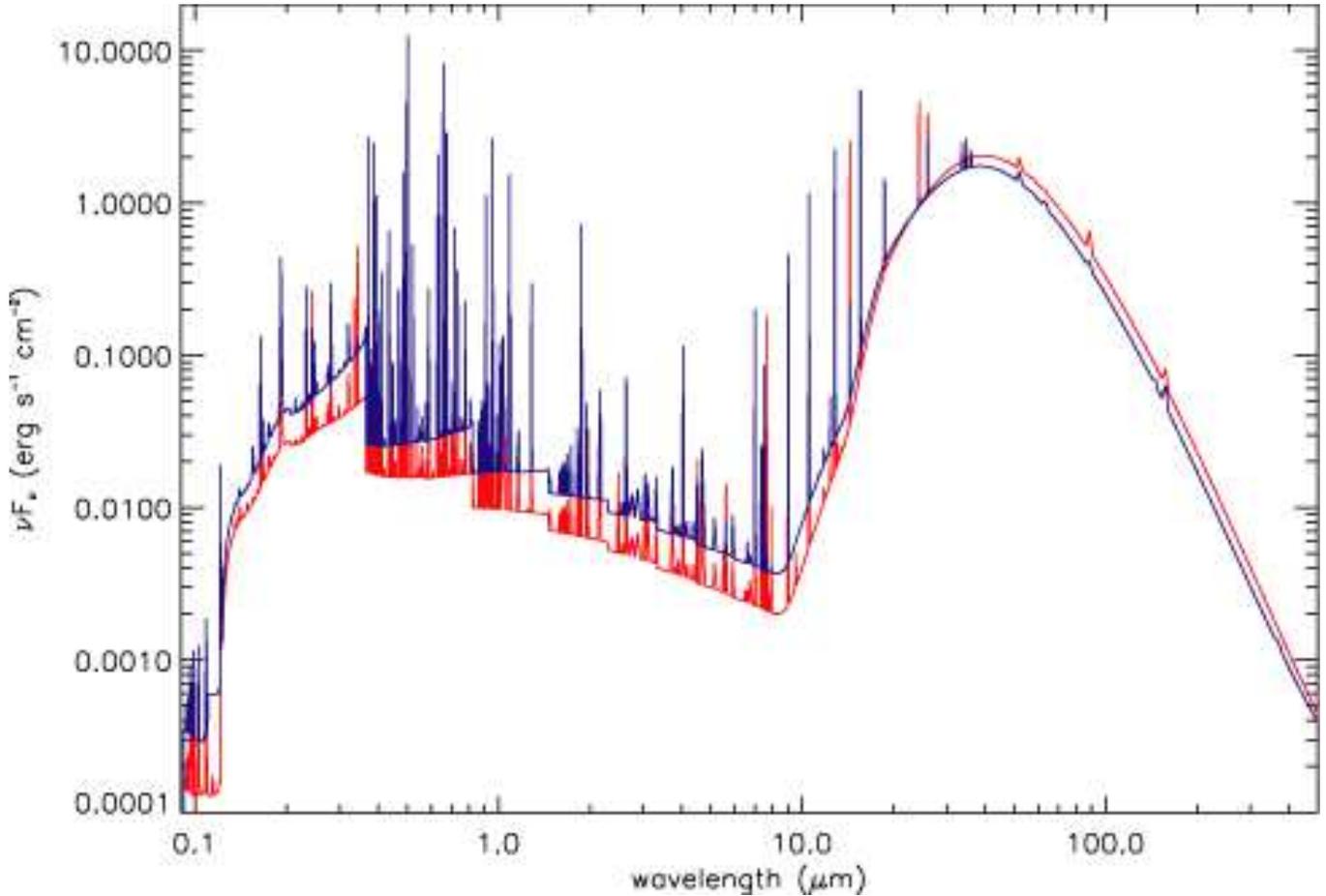}
\caption{A comparison of two models having the same input ionizing 
flux density but differing total pressures of $P_{\rm tot}/k\simeq 
10^7$ K \pccm\ ({\it red curve}) and $P_{\rm tot}/k\simeq 10^8$ 
K\pccm\ ({\it blue curve}) demonstrating the 
differences due to ionization parameter and pressure.
} 
\label{fig:sameI}
\end{figure*}

The effect of ionization parameter is more clearly seen in figure 
\ref{fig:sameI}, which shows two models with the same input ionizing 
intensity, but different total pressures, giving different 
ionizing parameters. The higher ionization parameter model (with 
$P_{\rm tot}/k\simeq 10^7$ K \pccm) shows a stronger IR 
dust feature relative to the lower ionization parameter model 
($P_{\rm tot}/k\simeq 10^8$ K \pccm). The dominance of dust 
absorption is most obvious however in the mid-IR to the optical-UV, where the 
log-scale emphasizes the separation in the models. This arises
predominantly due to the loss of ionizing photons to dust in the
higher $\Xi$ model which results in a
weaker free-free and free-bound continuum. This increased absorption
also produces  
more prominent FUV absorption.

\begin{figure*}[htp]
\includegraphics[width=\hsize]{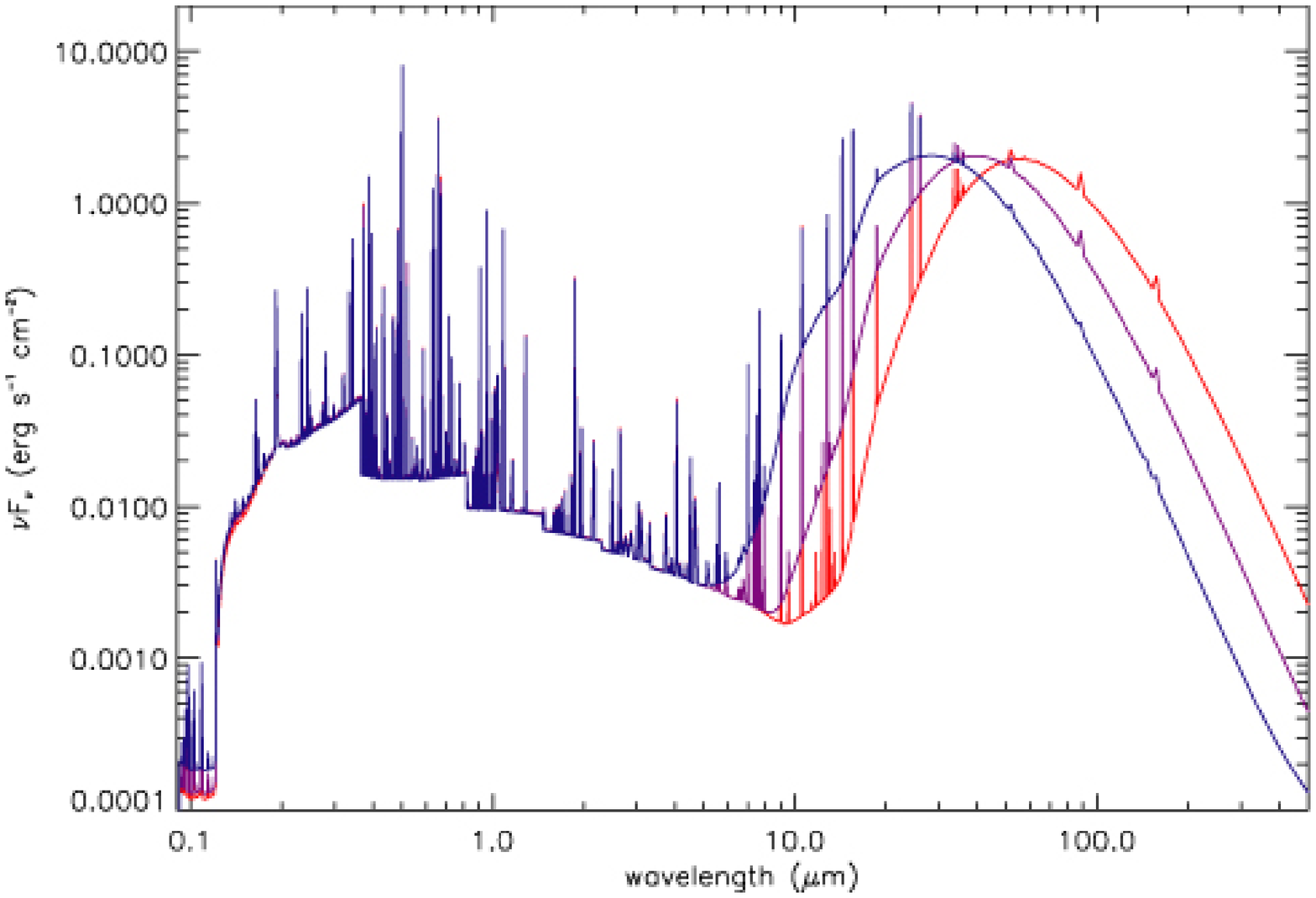}
\caption{A comparison of three model SEDs with the same ionization 
parameter,  $\log(\tilde{U}_{0})=-1.883$. The models have been scaled 
to remove the flux difference
due to the different input flux densities (see text).  As pressure 
decreases, the peak of the IR dust emission moves to longer
wavelengths.
}\label{fig:sameU}
\end{figure*}

To demonstrate the relative features that arise due to pressure
changes, figure \ref{fig:sameU} shows three model SEDs with the same initial 
ionization parameter ($\Xi=-0.66,\log(\tilde{U}_{0})=-1.883$) but with 
different $P_{\rm tot}$. To account for the different input flux 
densities, the models have been scaled by 10.0, 1.0, and 0.1 for the 
models of $P_{\rm tot}/k\simeq 10^6,~\simeq 10^7$ and 
$\simeq 10^8$ K\pccm\ respectively. 

Since they have the same 
ionization parameter, the gaseous line and continuum emission spectra 
of the models are very similar. However,  the progression of the dust 
IR re-emission bump feature to shorter wavelengths as we increase in 
$P_{\rm tot}$ is very marked. This progression can be explained by 
the increasing flux density experienced by the NLR cloud as the 
pressure increases to maintain the constant ionization parameter. 
The increasing incident flux density leads to higher 
average grain temperatures in the absorption region and thus stronger 
short wavelength emission. 

The UV spectra show that the same optical depth characterizes all
three models. With the same fraction of the absorbed UV light heating
the dust, the relative flux emitted by the dust in all three models is
the same, as can be determined from figure \ref{fig:sameU}.

When the models are allowed to continue to greater optical depths, 
the lower flux densities mean that the dust emits at longer
wavelengths. 
However, once the UV flux from 
the source is lower than the IR peak on a $\nu F_{\nu}$ plot, we don't
get much more  
far-IR emission, rather the dust peak merely becomes a little wider (see next
section). Thus the dust is hotter at all optical depths for a greater incident
flux density.  

Of special note is the presence of 
silicate emission features in the models with the highest incident
flux densities. This indicates that in the most luminous AGN or most
nuclear  narrow line regions, such emission is possible.

\begin{figure*}[htp]
\includegraphics[width=\hsize]{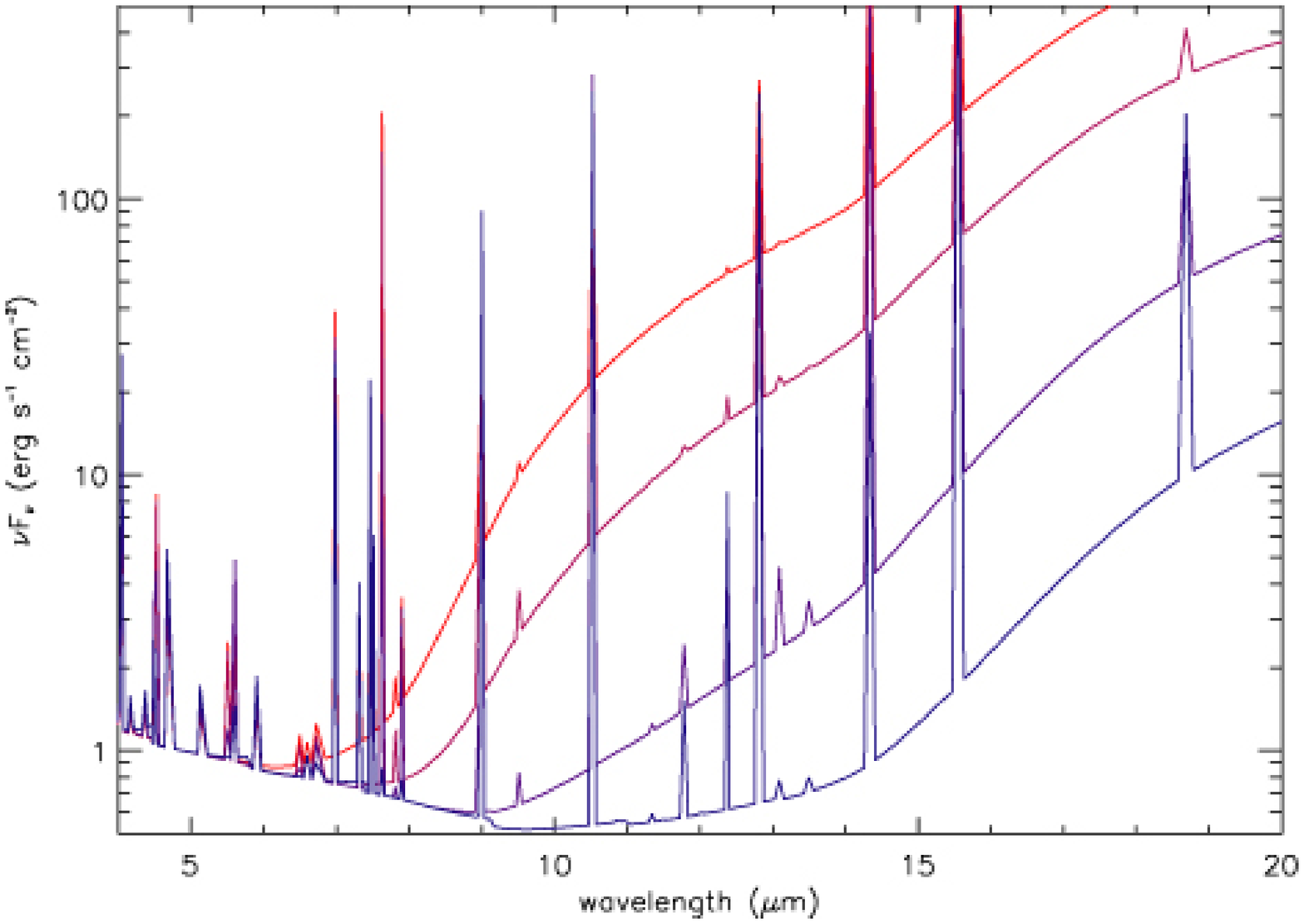}
\caption{The Mid-IR spectra of the $P_{\rm tot}/k\simeq10^7$ K \pccm\
NLR models from \ref{fig:1e7}. The models are shown here as $\log
(\nu F_{\nu})$ and are scaled to 1 at 5 \mum. The 10 \mum\ feature can be
seen as a slight curve in the higher ionization models (top
curves).}\label{fig:MidIR} 
\end{figure*}

Figure \ref{fig:MidIR} zooms in on the mid-IR region of the $P_{\rm
tot}/k\simeq10^7$ K \pccm SEDs of figure \ref{fig:1e7}. This figure reveals
in more detail the silicate features, especially that at 10 \mum. The
18 \mum\ feature is less visible due to the shape of the IR continuum
and its relative weakness. The figure
also demonstrates more clearly the increasing maximum temperature
achieved with higher incident fluxes, with the decreasing wavelength
or the IR feature edge.

\subsection{The Effect of Column Depth}

One final parameter to consider in the determination of NLR dust 
emission is the total column density of the narrow line region. The 
previous models were all calculated to a total column of 
$N($\hi$)=10^{21.5}$\pscm, such that they were predominantly 
ionization bounded. The column density affects the final line 
emission spectrum from the models as we either encompass (with 
increasing column density) or truncate (with lower column density) 
the various emission regions for each species.

\begin{figure*}[htp]
\includegraphics[width=\hsize]{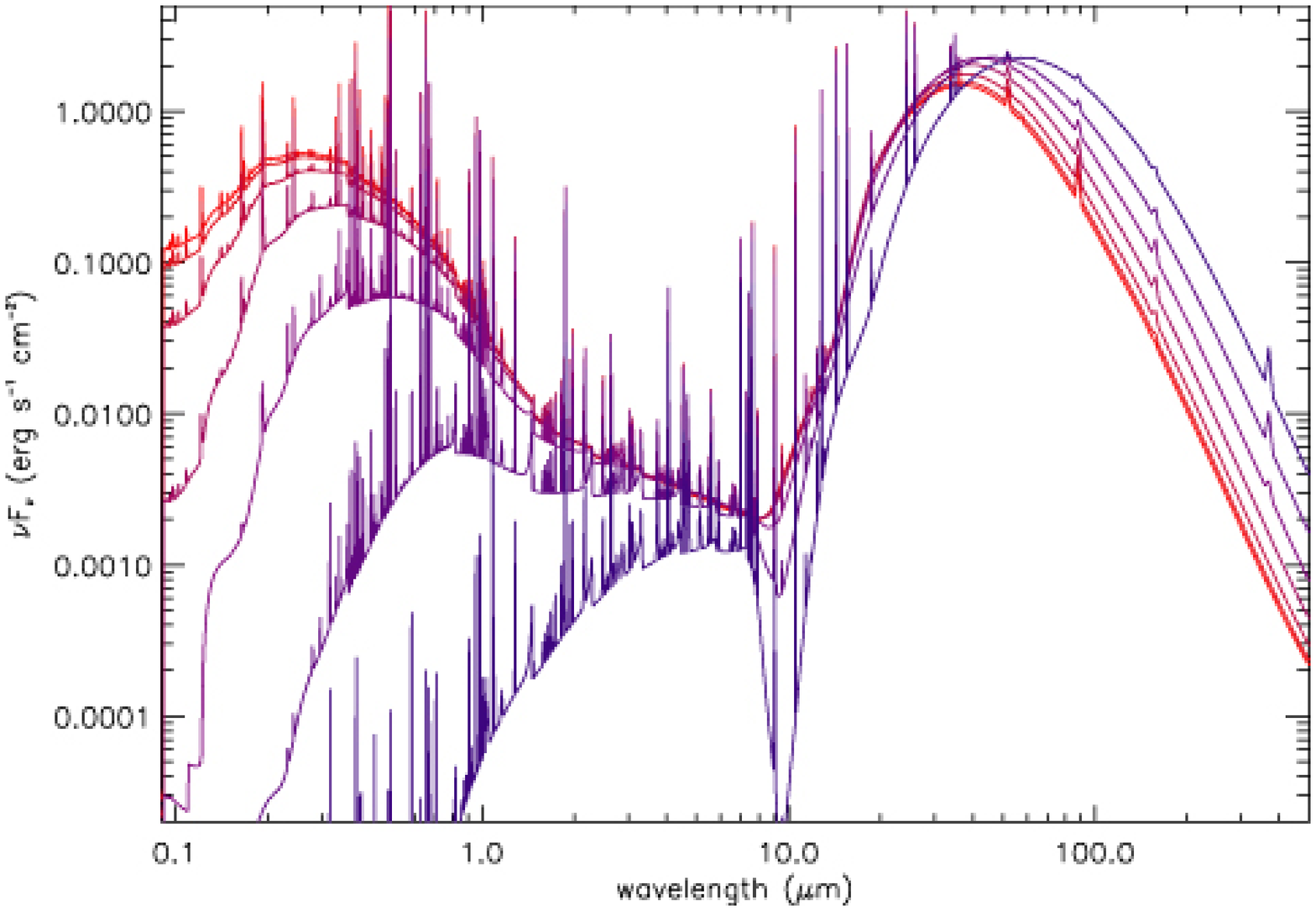}
\caption{Spectral energy distributions (SEDs) for NLR models with $P_{\rm
tot}/k=\simeq 10^7$ K \pccm, $\Xi=-0.66$, and
varying in total \hi\ column. The models increase in \hi column depth
from top to bottom (in the optical) from $10^{20.0}$ \pscm\ to
$10^{23.0}$ \pscm\ in steps of 0.5 dex. Note that the ionizing source
is included to demonstrate the total absorption.
}\label{fig:col}
\end{figure*}

As far as the dust is 
concerned, the greater the column density, the greater the optical 
depth of dust. At large column densities, most of the UV radiation 
field has been absorbed, and therefore the dust is at lower 
temperatures and emits weakly at longer wavelengths.  Figure 
\ref{fig:col} shows the effect of increasing the final column depth 
in a model with pressure of $P_{\rm tot}/k \simeq 10^7$ K\pccm\ 
and  ionization parameter of $\Xi=-0.66$. The column 
density is increased from $10^{20.0}$\pscm\ to $10^{23.0}$ \pscm\ in 
steps of 0.5 dex.

As expected, the low \hi\ column depth models show less extinction by 
dust and have a narrower and hotter dust emission bump. The 
highest column depth models have most of their optical absorbed by dust 
and show a much broader and stronger dust emission feature. The 
emission from the
coolest dust arises predominantly from the largest optical depths in
the models. This emission 
comes from the absorption of the low flux density visible and near-IR
photons. 

The 
spectrum in the 10-30 \mum\  is almost invariant with increasing 
optical depth until absorption becomes important, especially in the
strong 10 \mum Silicate feature. This is because the flux here is entirely 
contributed by the hottest dust in the ionized gas or just behind the 
ionization front.  
Strictly speaking, as these models are 1-D the results become 
unrealistic for high column densities. More realistic 3-D 
treatments of the dust radiative transfer is required in these cases,
especially as we are most likely viewing the NLR at
various angles and do not in general have a direct line of sight through the
obscuring material to the emitting edge of the NLR clouds.

\section{Discussion}

\subsection{The Shape of the NLR IR spectrum}

It is only in recent times that the resolution of infrared telescopes
has been high enough to begin to spatially resolve the emission from
narrow line region, and only in the nearest of active galaxies
\citep{Bock00,Galliano05}.
In all other active galaxies it is 
difficult to distinguish the relative contributions of starbursts and 
of AGN in the nuclear regions. Nonetheless, 
it is important to try to do this, in order to understand the 
energetics of the nucleus as a whole, as well as the proposed AGN - starburst 
connection. 

A comparison of the model spectra shown in the previous section
with the recent high spatial resolution
spectroscopy of NGC 1068 by \citet{Mason06} reveal several
similarities. The observed  NGC 1068 NLR spectra 
\citep[Fig. 2,]{Mason06}, show that the wavelength of the spectral
break between the nebula and dust continuum decreases with closer to
the nucleus. This indicates hotter dust temperatures with increasing
incident flux, exactly as in the models.  The more nuclear spectra
also show increasing silicate absorption due to increasing optical
depth. This probably precludes the detection of any silicate emission
from the inner NLR or nucleus itself.

However the observed NLR spectra also reveal much flatter Mid-IR spectral
slopes than obtained in most of the NLR models. The observations also
indicate that the dust is hotter than found in our simple models
discussed here. This suggests that there is some component of dust
that the models are missing. This is supported by the measurements of
\citet{Gratadour06} who find constant high temperatures of $\sim$500 K
out to 70pc from the nucleus of NGC 1068, where the NLR lies. This is
much hotter than can be predicted from simple NLR models
\citep{Barvainis87}.

The NLR of AGN most likely
obtains a greater range of density and ionization than seen in our
single cloud models which would soften this slope with the
combination of both high and low temperature dust. 
More importantly it would also contain a mix of density-bounded
(low \hii\ column density) and ionization bounded (low final \hii\
fraction, as in the models presented here) clouds. This mix would
allow the contribution of hot dust emission without significant
absorption, as well as the stronger cool dust emission. 
This can also be accomplished though the combination of low covering
factor hot dust, and large covering factor cool dust in a 3-D
geometry.

Smaller dust grains may provide another possible way for the models to
reach higher temperatures. Currently, the
average dust temperature of the front of the ionized cloud in the
higher ionization  $P/k=10^7$ is $\sim$250 K, and cooler temperatures
as we step into the cloud. The smallest grains in the model, while
having an average temperature lower than this, occasionally reach much
higher temperatures due to stochastic effects, and provide the hottest
IR. They also tend to dominate the emission due to the grain
distribution and larger surface area to mass. Thus, extending the
distribution to even smaller sizes increases the hottest dust
emission, although it does not assist greatly in flattening the Mid-IR
slope. 

Another possible explanation for the difference between the model and
observed slope is alternative heating mechanisms not considered within
these models, such as shocks. Although collisional heating is taken
account of in the models, shock heating and destruction of dust grains
can result in a much hotter IR spectrum, as considered in detail by
\citet{Contini04}. Even though emission line ratios indicate that
photoionization is the dominant ionization mechanism of the NLR in
most AGN, the correlation of jets with the NLR and high gas
temperatures and velocities observed within the NLR all indicate the
shocks play a part in this region \citep[see eg][]{Prieto05}. The
combination of shock and 
radiation heating of dust may explain the observed slope and dust
temperatures observed in the NLR of NGC 1068 and other AGN.

\subsection{The IR flux from the NLR}

While the contribution of the nucleus to the 
IR emission will presumably scale with the luminosity  of the nucleus 
itself, the contribution of the NLR to the total IR emission of a 
galaxy will depend upon both the AGN luminosity and the NLR covering 
factor. The models presented here allow, in principle, an estimation 
of the contribution of the NLR to the far-IR continuum. This 
exploits the coupling that exists between far-IR dust emission of the 
NLR and the emission lines generated within the NLR.

\begin{figure}[htp]
\includegraphics[width=\hsize]{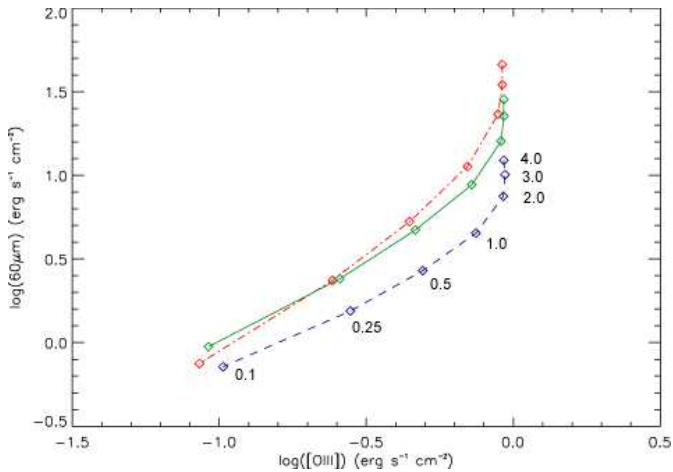}
\caption{The variation of the [\oiii] 500.7 nm flux against the IRAS 
60 \mum\ flux for the NLR models. Shown are the $P_{\rm 
tot}/k\simeq 10^6$ ({\it dash dot curve}), $\simeq 10^7$ ({\it 
solid curve}), and $\simeq 10^8$ ({\it dashed curve}), with the 
incident ionizing flux increasing from bottom to top along each curve as
labelled. 
To remove the effect of source intensity variation the models have 
been scaled by 10.0, 1.0, 0.1 for the $P_{\rm tot}/k\simeq
10^6,~\simeq 10^7$ and $\simeq 10^8$ K\pccm\ models respectively.}
\label{fig:OIII60}
\end{figure}

Figure \ref{fig:OIII60} shows the relationship between the strongest 
emission line indicator for NLRs and AGN, the [\oiii] 500.7 nm line, 
and the IRAS 60 \mum\ flux (broadly equivalent to the  the Spitzer 
MIPS 70 \mum\ band). Note that the models have been scaled to remove 
the effect of source intensity, as an increase in source luminosity 
will result in both higher [\oiii] and 60 \mum\ luminosity, when 
other parameters are kept constant.  The ratio of the two fluxes 
shows some variation due to pressure, but more due to changing 
ionization parameter. At  low ionization parameter, $I_{[OIII]} \sim 
0.2\times I_{60 \mum}$, but at the highest ionization parameter 
$I_{[OIII]} \sim 0.04\times I_{60 \mum}$.  However both pressure and 
ionization parameter can be estimated through the measurement of 
other emission lines such as [\sii] and H$\alpha$ \citep[see 
e.g.][]{Groves04b}, reducing the possible range of the 
[\oiii]/60\mum\ ratio.

A potential problem in comparing the two fluxes is that the [\oiii] 
line may be heavily extinguished by dust along some lines of sight. 
In these cases, the theoretical ratio of the  60 \mum\ flux to  the 
[\oiii] 500.7 nm line can be used to provide an upper estimate of the 
NLR luminosity. However, a better measure of the AGN luminosity 
absorbed by the NLR may be obtained using IR emission lines such as 
[\nev] 14.3 \mum\ or [\oiv] 25.8 \mum. 

The 60 \mum\ flux is also 
affected by the extent of the absorbing column of the NLR, with 
greater
columns giving greater IR flux. A better measure would be to use the 
25 \mum\ band as
it remains much less affected by the column depth, at least up to the 
point where Silicate absorption becomes important ($\sim 
N($\hi$)\gapprox10^{22.5}$, figure \ref{fig:col}). At these columns 
the IR emission lines will likewise be affected by dust
extinction. The 25\mum\ band is also where the NLR flux is most likely
to dominate, lying between the 12\mum\ band, dominated by hot
torus-like dust, and the FIR bands (60\mum\ and 100\mum\ bands) where
star formation is most likely to dominate.

\begin{figure}[htp]
\includegraphics[width=\hsize]{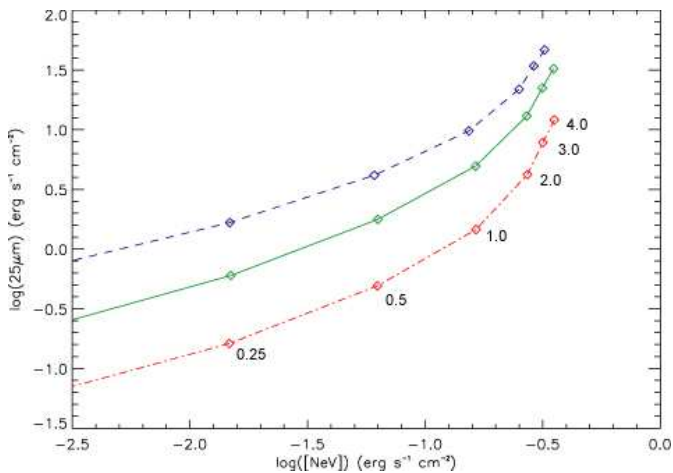}
\caption{The variation of the [\nev] 14.3 \mum\ flux against the IRAS
25 \mum\ flux for the NLR models. Curves and scaling are as in figure 
\ref{fig:OIII60}
}\label{fig:NeV25}
\end{figure}

The [\nev] 14.3 \mum\ flux is plotted against the IRAS 25 \mum\ flux 
in figure \ref{fig:NeV25}. The 25 \mum\ continuum shows a much 
greater sensitivity to pressure due to the changing temperature of 
the hottest dust. The [\nev] line is much more sensitive to the 
ionization parameter than the [\oiii] line. Together, these effects 
are expected to produce a scatter of ten in the correlation between 
the [\nev] 14.3 \mum\ and the IRAS 25 \mum\ fluxes.

\begin{figure}[htp]
\includegraphics[width=\hsize]{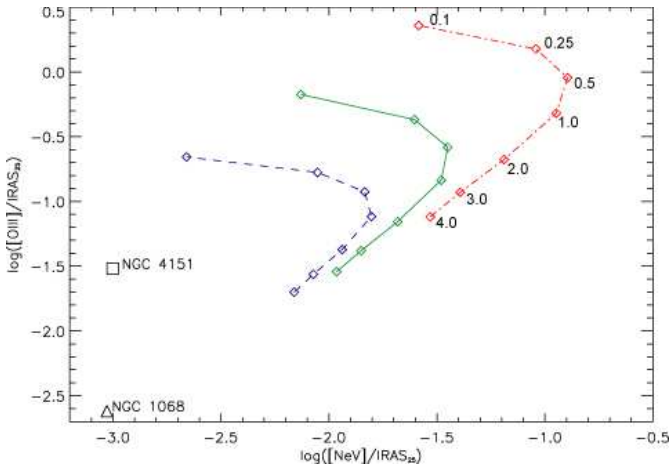}
\caption{Diagnostic [\nev] 14.3 \mum/IRAS 25 \mum\ versus [\oiii] 500.7
nm/IRAS 25 \mum\ ratio diagram. Each curve shows a different pressure,
decreasing from left to right. The direction of increasing ionization
parameter for the curves is indicated by the values of $I_{0}$ marked
next to the $P/k=10^6$ model. The symbols show the observed ratios of
NGC 1068 and NGC 4151 (as marked), with measurement errors dominated
by line flux calibration of $\sim$20\%. 
Data from
\citet{Alexander99,Alexander00,Sturm02}. 
}\label{fig:NeV25ratio}
\end{figure}

If we combine these lines we can get a good diagnostic for both
the fraction of infrared emission arising from the NLR and the
extinction experienced in the optical. Figure \ref{fig:NeV25ratio}
demonstrates this with the ratio of [\nev] 14.3 \mum/25 \mum\ versus
[\oiii] 500.7 nm/25 \mum. The first ratio gives an estimate of the NLR
contribution to the 25 \mum\ flux. If the observed value of this ratio
is less than that given by the models, then it suggests that not all
of the 25 \mum\ flux arises from the NLR. However, the 25 \mum\ flux
is strongly affected by the pressure of the models, or more precisely
the incident flux density at a given ionization parameter. At a given
ionization parameter, as the
 pressure and incident flux are increased, the [\nev] 14.3\mum\
line remains approximately constant, at least for the lower
densities. 
However with the increase in incident flux, the 25\mum\ flux
is increased, decreasing the [\nev]/25 \mum\ ratio. 
Thus the pressure or incident flux
must be determined to give an accurate estimate of this
ratio. In the IR this can be done using the density
sensitive ratio of [\nev] 14.3/24.3 \mum\ (see figure \ref{fig:dens}), 
thus connecting directly with this diagram.

The [\oiii]/25\mum\ ratio can also provide an estimate for the NLR
contribution to the IR, but in addition gives an indication of the
obscuration of the NLR. As the NLR becomes more heavily obscured this
ratio will be reduced by the loss of the [\oiii] line. 

This can be seen in the two observational points in this diagram. NGC
1068 and NGC 4151 are two of the closest and most studied active
galaxies. Respectively these are quite often used as the
representative Type-2 and Type 1 AGN, corresponding to strongly obscured,
edge-on, and lightly obscured, face-on AGN in the unified model. 

NGC 4151 appears to the lower left of the region occupied by the models.
The observed mid-IR [\nev] and [\siii] lines give a density of $\sim
10^3$\pccm\ (figure \ref{fig:dens}), indicating the ratios are
less by 1.0 to 1.5 dex. This suggests that the NLR contributes
only about 3 -- 10\%
to the total 25\mum\ emission, depending on the ionization state of the gas.

NGC 1068 has a similar
[\nev]/25 \mum\ ratio to NGC 4151, [\oiii]/25\mum\ ratio of NGC
1068 appears less by $\sim 1$ dex compared to NGC 4151. While the
higher density of NGC 1068 indicated by the mid-IR lines ($\sim 10^4$ \pccm,
figure \ref{fig:dens}) and a differences in the average ionization
parameter can possible explain this difference,  some part of the
difference may also arise from the attenuation of the [\oiii]
flux. This would fit within the Type-1 and 2 paradigm and is supported
by the NGC 1068 emission line asymmetries \citep{Lutz00} and estimates
of low reddening in NGC 4151 \citep{Kriss95}.

Assuming a higher density of $\sim 10^4$ \pccm\ and higher ionization
parameter figure \ref{fig:NeV25ratio} suggests that the NLR of NGC
1068 has a similar contribution of
about 3 -- 10\% to the total 25\mum\ emission.

Obviously this is a low contribution, and means that there are other
effects not considered within the simple models
presented here. 

One possible consideration is the effect of aperture. The [\nev]
emission line arises only from the AGN, and likewise the [\oiii] is
dominated by the AGN contribution. By contrast, the IRAS aperture is large
and can include the contribution of star formation in
both NGC 1068 and NGC 4151. Recent high spatial resolution
observations suggest that this contribution is not large,
approximately 10\% or less \citep{Bock00,Radomski03}. This does not
preclude any nuclear star formation contribution within the $\sim
4\arcsec$ resolution of these observations.

Another possibility is that the NLR of these objects is a different
metallicity to that considered here. Most AGN NLR appear to be best
fit by metallicities of about $\sim 2$Z$_{\odot}$
\citep{Groves04b}. An increase in metallicity would increase the dust
absorption relative to the gas and hence the IR  emission. The
increase relative to the [\nev] line is more uncertain as it is not
depleted onto dust, but will be weaker due to the lower gas
temperature in higher metallicity gas.
In addition to metallicity differences, an increase of the dust
to gas ratio through increased depletion could also increase the
contribution of the NLR to the total IR, though it must be balanced
with the line emission.

However the most likely possibility is that there are other components
of IR emission not contained within our simple models. The most
obvious of these is the putative torus, or more precisely, the equatorial
dusty gas that is needed to explain the unified model, and probably
associated with the accreting gas. This gas must be of a high enough
density that it has no associated forbidden emission lines, such as
[\oiii], are emitted and close enough to the nucleus that hot dust
temperatures are achieved.  

In addition to this there may also be dusty coronal gas, within the
NLR that is so hot that species such as \oiii\ have been
ionized. While dust may have only a short lifetime within this region
due to sputtering, the covering fraction of this gas may be large
enough to contribute to the total IR emission.

Connected with this is the possibility of shock dust heating. As
mentioned in the previous section, shocks are known to exist within
AGN, and may contribute to the total IR emission. \citet{Contini04}
were able to explain the IR spectrum and flux of several AGN using
shocks alone, suggesting that they may be a viable contribution to the
total IR. 

One other source of dust heating may be nuclear star
formation. Nuclear starformation is known to exist in several AGN
through the presence of PAH emission features \citep[eg]{Imanishi04},
and will contribute to the total IR emission without contributing to
the [\nev] emission. Nuclear star formation may contribute to other
emission lines however such as [\oiii], though the optical-UV lines
may be heavily attenuated. 

\subsection{Mid-IR Emission Line Diagnostics}

Any nuclear star formation will also produce emission lines, although not 
of as high ionization potential as \nev. A potential means to separate 
the contribution to the IR originating from AGN and that coming from 
circum-nuclear star formation is therefore to examine IR line flux 
ratios.  Diagnostic ratios from ISO, such as that of  
polycyclic aromatic hydrocarbon (PAH) features against the hot dust 
continuum or strong emission lines
\citep{Lutz98,Genzel98,Lutz03}, have also been considered for this 
purpose. Here, we restrict our 
examination to the IR emission lines found in both starforming 
galaxies and AGN. 

The choice of IR emission line diagnostics is
similar to that in the optical regime. The chosen emission line ratios must
be strong, and generally obtainable within the same spectral
range. One important difference is that extinction and reddening of
the lines is no longer a strong criterion for line choice. 
Several good diagnostic
ratios have previously been suggested by \citet{Alexander99} and
\citet{Sturm02}.  

\begin{figure}[htp]
\includegraphics[width=\hsize]{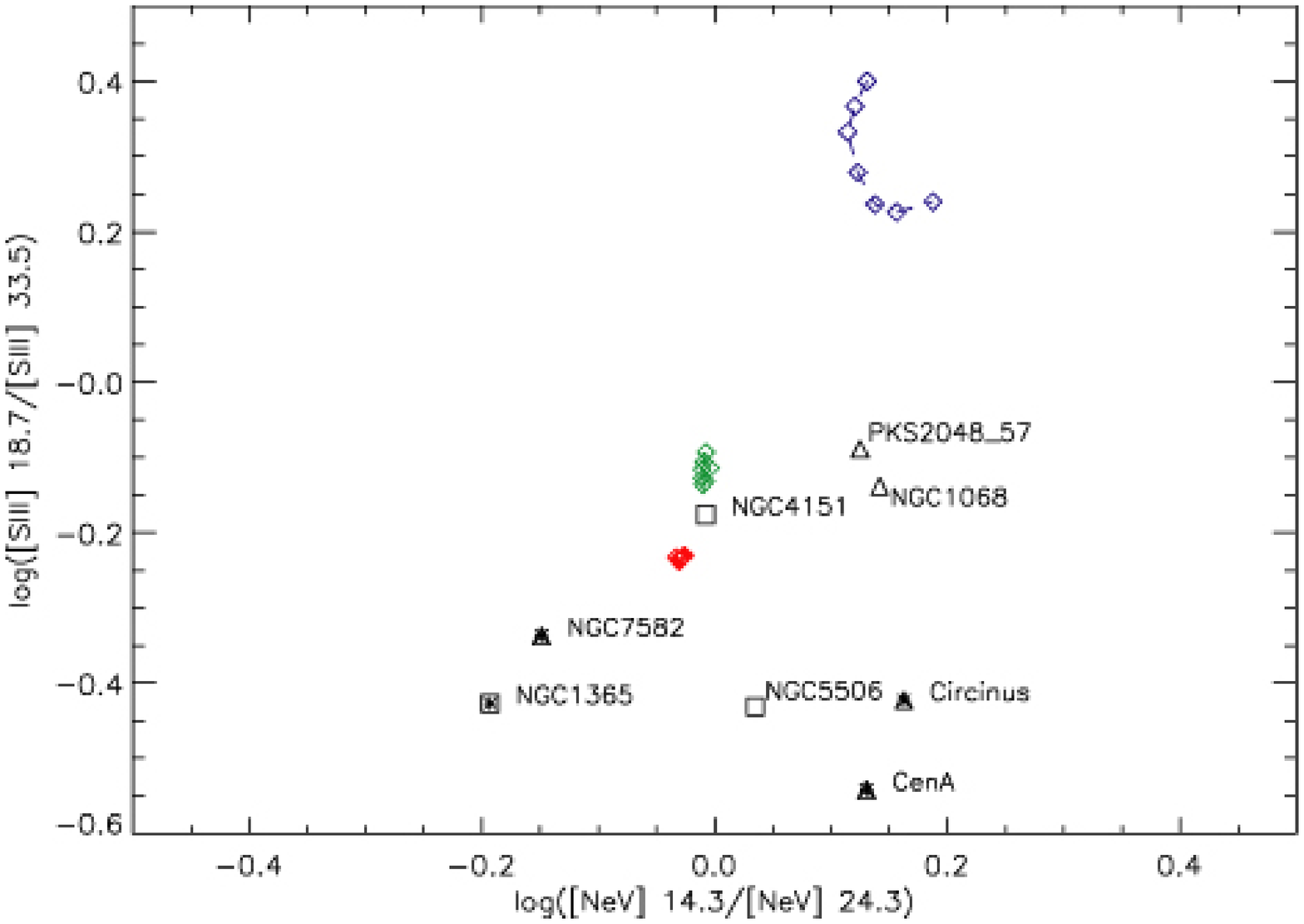}
\caption{Density sensitive [\nev] 14.3 \mum/[\nev] 24.3 \mum\ versus [\siii]
18.7 \mum/[\siii] 33.5 \mum\ ratio diagram. Each curve shows a
different pressure, 
increasing from left to right, as labelled. 
The symbols show the observed line ratios of nearby AGN (as labelled) from the
ISO-SWS survey of \citet{Sturm02}. Triangles represent Seyfert 2
galaxies, while squares are Seyfert 1s, with the star filled objects
representing those with associated star formation within the
aperture. The average ratio error of the AGN is $\sim$28\%.
}\label{fig:dens}
\end{figure}

Two such ratios discussed in previous works are the density sensitive line
ratios [\nev] 14.3/[\nev] 24.3 \mum, and [\siii] 18.7/[\siii] 33.5
\mum, as shown in figure \ref{fig:dens} \citep[see][for a description
of density sensitivity]{Alexander99}. These ratios can help remove
some of the uncertainty seen in the previous diagrams. In this figure
(and figures \ref{fig:dens} and \ref{fig:OVIiras}) the observational
data is from \citet{Sturm02}, with relevant discussions on the precision of the
data found within.

Both ratios consist of strong
IR lines and remove direct  ionization and metallicity effects. 
As [\nev] is such a high ionization species, it arises purely from the
NLR, and in particular gives an indication of the densities  in the
hot, highly ionized regions. The [\siii] ratio gives an idea of the
densities in the cooler parts of the NLR, but can also be contaminated
by emission from nuclear star formation. 

Such a contribution may explain the offset between the observed AGN
and the models, where the [\nev] ratio suggests a higher density than
the [\siii] ratio for several objects. However, no coincidence with
associated starbursts is seen (starred objects). Another possibility
 is that the region of the NLR where the [\siii] lines arise
is at a lower density than the [\nev] region. Like previous work, we
use the [\nev] ratio for our estimated density.

To separate the nuclear starformation from AGN activity,
the line ratio diagnostics generally depend upon lines that have either an
ionization beyond that found in star forming regions, such as [\nev]
14.3\mum, or are sensitive to the extended partially ionized region
found around AGN, like [\silii] 34 \mum.  Such examples are shown in
\citet{Sturm02}. The problem with the high
ionization potential  lines is that they are very sensitive to the
ionization conditions of the nebulae. Similarly, the lines arising in
the partially ionized region are generally matter bounded, and are
sensitive to the total column depth of the emitting region. 

\begin{figure}[htp]
\includegraphics[width=\hsize]{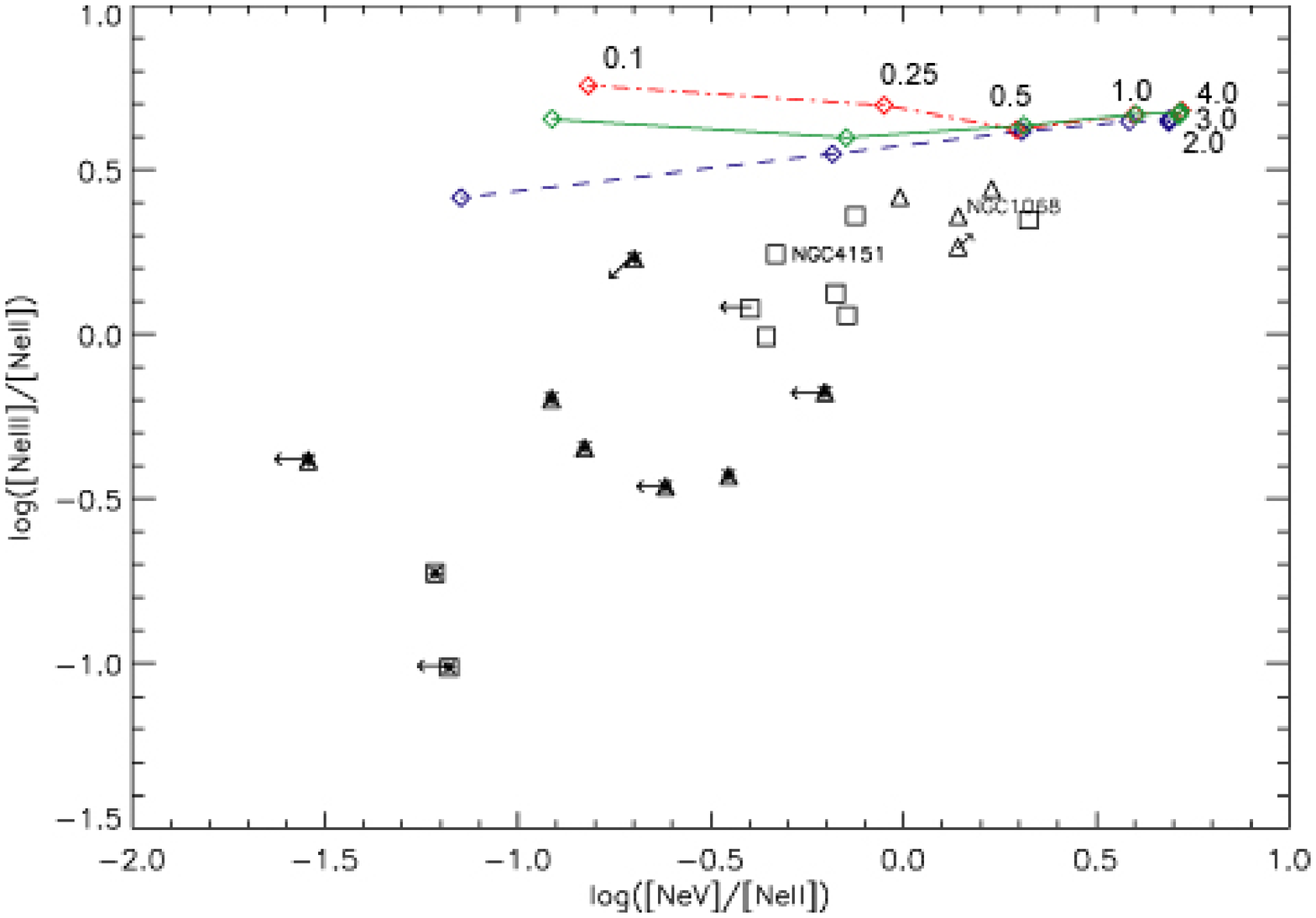}
\caption{Ionization source diagnostic diagram of  [\nev] 14.3
\mum/[\neii] 12.8 \mum\ versus [\neiii] 15.5 \mum/[\neii] 12.8 \mum. 
The direction of increasing ionization
parameter for the model curves is indicated by the values of $I_{0}$
at each model point. The observed AGN ratios are marked as in figure
\ref{fig:dens}, with the same average measurement errors of 28\%. Any
contribution of starformation will move objects to the lower left of
the diagram.
}\label{fig:Neratio}
\end{figure}

Figure \ref{fig:Neratio} shows a diagram which partly avoids these
problems. It uses three close proximity, strong neon emission lines;
[\neii] 12.8 \mum, [\nev] 14.3 \mum\ and [\neiii] 15.5 \mum. The
[\nev]/[\neii] ratio is a strong indicator of the AGN contribution, as
a nuclear starburst (SB) will only contribute to the [\neii] line and thus only
weakens this ratio. Typical NLR values for this ratio are 3 --
10. However as seen from the model curves in figure \ref{fig:Neratio}
this ratio is very sensitive to the ionization conditions, thus is
degenerate between AGN-SB mixing and ionization. The second neon
ratio, [\neiii]/[\neii], helps to remove this degeneracy. As shown by
the model curves, it is much less sensitive to the ionization state of
the gas. While starbursts produce both these emission lines, the
ratio is much lower than found in AGN (depending upon the metallicity
of the SB).

The usefulness of this diagram can be seen in the distribution of the
observed AGN, with all objects with a known starburst component lying
to the lower left of the diagram. A possible empirical dividing line
for starburst affected AGN then suggested by this diagram is
\begin{equation}
[\mathrm{NeIII}]/[\mathrm{NeII}] < 0.9 ([\mathrm{NeV}]/[\mathrm{NeII}]) - 0.35.
\end{equation}

A more theoretical mixing diagram will be presented in future work. 

One possible source of error in this diagram in stronger starbursting
galaxies, is the proximity of the neon emission lines to the PAH
emission bands, which may make measurement of these lines difficult
\citep[see e.g.][]{Weedman05}.

\subsection{Distinguishing the NLR and Dusty Torus in IR Emission}

While the line diagnostic ratio diagrams are a good way to separate
the contribution of starformation and AGN to a spectrum, separating
the contribution of the NLR and the putative inner, dusty torus is
more difficult. The only possible way is to use line--continuum
ratios, either as equivalent widths or line--band ratios. 
Figure \ref{fig:NeV25ratio} shows one possible diagnostic diagram,
though this requires
the subtraction of any star formation contribution to the 25 \mum\
flux to properly determine the torus contribution. 

\begin{figure}[htp]
\includegraphics[width=\hsize]{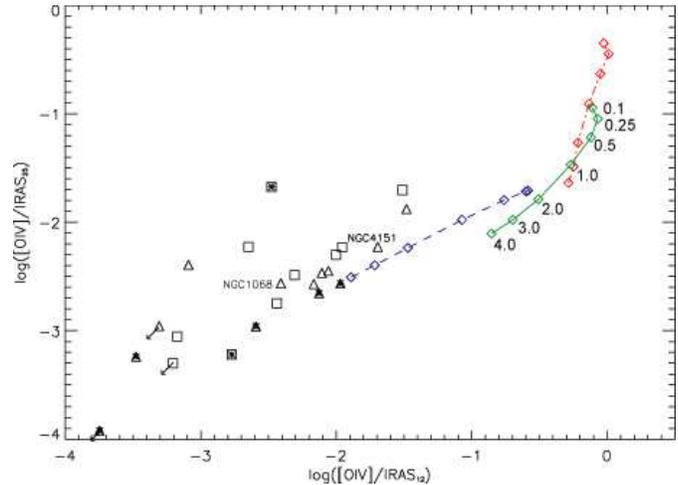}
\caption{IR line--continuum diagnostic diagram of [\oiv] 25.9 \mum/
IRAS 12\mum\ band versus [\oiv] 25.9 \mum/
IRAS 25\mum\ band. Observed AGN ratios are marked
as in figure \ref{fig:dens}. The IR contribution from a dusty torus
will move objects lower and strongly to the left. Averagee measurement errors
are approximately 20\% \citep[see][]{Sturm02}.
}\label{fig:OVIiras}
\end{figure}

A possible
alternative to this is shown in figure \ref{fig:OVIiras}. It uses a
strong, high ionization emission line, in this case [\oiv] 25.9 \mum, against
the 25 \mum\ and 12 \mum\ IRAS fluxes. This diagram is similar to the
IRAS colour--colour diagrams \citep[see eg][]{Dopita98}, but includes
the Mid-IR emission line to account for the IR emission from the NLR. 
We expect the dusty torus to be
significantly hotter than the NLR, and thus show stronger emission at
shorter wavelengths. Therefore the emission at 12 \mum\
should be less correlated with the [\oiv] emission line than the 25
\mum\ flux.

The observations support this, with the observed AGN appearing 0.5 to
1 dex lower than the models in [\oiv]/25 \mum, but approximately 1
to 2 dex lower in the [\oiv]/25 \mum\ ratio. This diagram also gives
an indication of a much greater contribution of the NLR to the IR
flux, and reveals the need to use several line ratios to get a good
indication of the NLR contribution. Depending on the density assumed,
the [\oiv]/25 \mum\ ratio indicates a NLR contribution of 10--50\% to
the 25\mum\ flux, and a weaker contribution to the 12\mum\ flux. 
Determining the exact contribution of the NLR 
to the total IR will
depend upon both the shape of the torus IR feature and the diffuse IR
emission.

A NLR contribution of 10--20\% to the mid-IR flux is not
unreasonable and fits in with observations. 
High resolution observations also show the dominance of the torus,
with sub-arcsecond images of NGC 1068 indicating that around 75\% of
the Mid-IR emission arises from the unresolved core where the torus
lies \citep{Bock00}. 

In terms of energetics, the contribution of the NLR depends critically to the
covering fraction of the NLR which is an unknown quantity. If dust is
ignored, a H$\beta$ line emission analysis of nearby Seyferts gives a
covering fraction of only $1\% $ --$4\% $ \citep{Netzer93}. However, 
when dust is
included it competes with Hydrogen for the ionizing photons
\citep{Dopita02}, and thus much greater covering fractions are
possible. Likewise, observations of nearby Seyfert galaxies appear to
support a covering fraction greater than a few percent
\citep{Netzer93}.

Another indicator that the NLR may have a large
covering fraction is the observations of UV or ``associated'' absorbers. These
absorption systems are believed to have a covering fraction $\gapprox$
50\% \citep{Crenshaw03}, and several observation indicate a possible
connection between these and the NLR \citep{Cecil02,Crenshaw05}. While
these systems do not have the column density observed in the NLR,
figure \ref{fig:col} shows that such systems can still contribute to
the IR emission if they contain dust. Thus a NLR contribution of $\sim
10\%$ to the total IR flux in AGN is not unfeasible.

As a last point, while these models do not represent in any form the
putative torus around the AGN from the unified model, this work can be
extended to this region, albeit without the consideration of
geometry. 
As we increase the density and incident flux
onto the models we start to reach the possible physical conditions
within the torus, as the appearance of silicate
emission in figure \ref{fig:MidIR} shows.

\section{Conclusion}

We have explored within this work a series of Dusty Narrow Line Region
(NLR) models, covering
a range of realistic pressures and ionization parameters,
with the aim of investigating the Infrared emission from dust in these
regions.

The presented spectral energy distributions (SEDs) for these models
reveal the possible range of the IR emission from the NLR, and the
correlation of this emission with the ionization parameter and the
intensity of the ionizing radiation. An
increase in either of these parameters will increase both the total IR
flux from the dust and the peak temperature of the dust.

Comparisons with observed NLR IR spectra show that simple 1-D photoionization
models are not enough to represent the complexity of the NLR. 
Combinations of different NLR clouds and NLR dynamics may be needed to
represent a more accurate model of the NLR IR emission.

While examining the spectral energy distributions of these models
gives an idea of both the overall emission from the NLR
of active galaxies, the greatest benefit of these models is that they give a
direct connection between the emission by dust and the line emission
from the photoionized gas.  A selection of correlations between line
and dust emission are presented here to give possible diagnostics for
the IR emission from AGN and galaxies. In addition we have explored
further possible IR emission line diagnostics to assist in the
determination of the NLR physical conditions, and to help distinguish
the contribution that starformation makes to the total spectrum.

Comparisons of the models with the nearby AGN, NGC 1068 and NGC 4151,
appear to show that only $\sim$10\% of the 25\mum\ emission arises from
the [\oiii] line emitting region. This means a significant fraction
must arise from either the putative torus, nuclear star formation,
coronal--shock excited region, or some combination thereof.

While the fraction the NLR contributes to the overall
infrared continuum of an active galaxy may be small, 
it is important to consider this component when
trying to understand the full spectral energy distribution of AGN. The
final aim will be to connect the emission from the NLR with what is
occurring in the central engine and thus connect the emission lines
with the total IR output of the AGN.

\begin{acknowledgements}
M. Dopita 
acknowledges the support of both the Australian National University 
and of the Australian Research Council (ARC) through his ARC 
Australian Federation Fellowship.  Dopita \& Sutherland 
recognize the financial support of the ARC through  Discovery project 
grant DP0208445.
\end{acknowledgements}

\end{document}